\documentclass[journal]{IEEEtran}
\usepackage{color}
\usepackage{graphicx}
\usepackage{epstopdf}
\usepackage{amssymb,amsmath}
\usepackage{amsthm}
\usepackage{cite}
\usepackage{relsize}
\usepackage{subfigure}

\bstctlcite{IEEEexample:BSTcontrol}
\newtheorem{proposition}{Proposition}

\newtheorem{lemma}{Lemma}
\newtheorem{corollary}{Corollary}
\theoremstyle{definition}

\newcommand\numberthis{\addtocounter{equation}{1}\tag{\theequation}}
\newcommand\blfootnote[1]{%
  \begingroup
  \renewcommand\thefootnote{}\footnote{#1}%
  \addtocounter{footnote}{-1}%
  \endgroup
}
\IEEEaftertitletext{\vspace{-2\baselineskip}}

\ifCLASSINFOpdf

\else

\fi

\begin{document}

\title{Grant-Free Massive NOMA:\\Outage Probability and Throughput}

\author{Rana~Abbas,~\IEEEmembership{Student Member,~IEEE,}~Mahyar~Shirvanimoghaddam,~\IEEEmembership{Member,~IEEE,}~Yonghui~Li,~\IEEEmembership{Senior Member,~IEEE,}~and~Branka~Vucetic,~\IEEEmembership{Fellow,~IEEE}}
\maketitle
\begin{abstract}
In this paper, we consider a massive uncoordinated non-orthogonal multiple access (NOMA) scheme where devices have strict latency requirements and no retransmission opportunities are available. Each device chooses a pilot sequence from a predetermined set as its signature and transmits its selected pilot and data simultaneously. A collision occurs when two or more devices choose the same pilot sequence. Collisions are treated as interference to the remaining received signals. We consider successive joint decoding (SJD) and successive interference cancellation (SIC) under a Rayleigh fading and path loss channel model. We first derive the expression for the outage probability for the case where devices transmit at the same fixed rate. Then, we derive the expression for the maximum achievable throughput for the case where devices transmit with rateless codes. Thus, their code rate is adaptive to the system conditions, i.e., load, received powers, and interference. Numerical results verify the accuracy of our analytical expressions. For low data rate transmissions, results show that SIC performs close to that of SJD in terms of outage probability for packet arrival rates up to 10 packets per slot. However, SJD can achieve almost double the throughput of SIC and is, thus, far more superior.
\end{abstract}
\section{Introduction}
\blfootnote{The authors R. Abbas, M. Shirvanimoghaddam, Y. Li and B. Vucetic are with the Center of Excellence in Telecommunications, School of Electrical and Information Engineering, The University of Sydney, NSW, Australia (email: \{rana.abbas; mahyar.shirvanimoghaddam; yonghui.li; branka.vucetic\}@sydney.edu.au).}
\blfootnote{This paper was submitted in part to PIMRC 2017.}
\IEEEPARstart{M}assive Machine Type Communication (mMTC) is expected to play an important role in future 5G networks \cite{mM2M2016bockelmann}. mMTC handles the connections between the massive number of machine type devices (MTDs) which are characterized by their short packets, and low computational and storage capabilities. Moreover, in most cases, MTDs have fixed locations and need to operate on a low power budget. Previous works have shown that coordinated access is not suitable for mMTC due to the large control overhead. The control overhead can be as large as the payloads themselves in a large network which is very inefficient \cite{durisi2016toward}. In contrast, uncoordinated access operates without or with minimal control overhead at the expense of collisions due to devices contending to access the channel.

One of the most common uncoordinated access schemes is the slotted ALOHA protocol. Many variants of this protocol have been proposed over the past decade \cite{IRSA,popovskiLetter,RA_mag} that aim at improving the system throughput. However, as these protocols are orthogonal in nature, i.e., transmissions take place over a set of non-overlapping resource units, the number of devices that can be supported is dependent on the number of available resource units. Moreover, from an information theoretic perspective, orthogonal multiple access has been shown to be strictly sub-optimal for short packet transmissions \cite{molavianjazi2014second}. The gap between the achievable sum-rate of orthogonal multiple access and the maximum sum-rate increases as the system load increases \cite{molavianjazi2014second}.

On the other hand, authors in \cite{molavianjazi2012random} showed that the maximum sum-rate is achievable through non-orthogonal multiple access (NOMA) and joint decoding. This is because NOMA allows multiple users to share the different resources, e.g., time, frequency, and space, either through power domain multiplexing or code domain multiplexing  \cite{noma2015dai,mM2M2016bockelmann,noma2016yuan}. Thus, unlike orthogonal multiple access, overloading is possible at the expense of increased processing complexity at the receiver \cite{noma2016yuan}. As mMTC communications are uplink-oriented, this complexity is at the access point (AP) and is, thus, acceptable. NOMA allows multiple users to share time and frequency resources in the same spatial layer via power domain or code domain multiplexing. In what follows, we highlight some of the recent works conducted on NOMA so far.
\subsection{Related Work}
For uplink coordinated NOMA, the problem of user scheduling, resource allocation, and power control has been widely investigated \cite{takeda2011enhanced,endo2012uplink,al2014uplink,haci2015novel,sung2016game,zhang2016uplink,sedtheetorn2016accurate,tabassum2017modeling,yang2017non}. Authors in \cite{haci2015novel} proposed a novel interference cancellation technique, called Triangular SIC, for asynchronous NOMA. Triangular SIC detects and exploits a priori information of all the overlapping symbols. Authors also presented the bit-error rate for their proposed technique. In \cite{yang2017non}, authors proposed a physical layer network coding scheme and cascade-computation decoding scheme for the case where a fixed number of users are transmitting over a fading channel. Their results show that cascade computation can outperform the iterative detection and decoding scheme. Authors in \cite{sedtheetorn2016accurate} considered a single-cell scenario where active users follow a Poisson Point Process (PPP). All the active users are known at the AP and are scheduled together. Authors derived the expression for the achievable spectrum efficiency for Nakagami fading. In \cite{tabassum2017modeling}, authors considered a multi-cell scenario. Using the theory of ordered statistics and Poisson Cluster Process, they characterized the rate coverage probability for perfect and imperfect SIC. Their results show that NOMA outperforms its counterpart orthogonal multiple access cluster in the cases of higher number of users per cell and higher target rate requirements.

For uncoordinated NOMA, the problem is that the set of active users as well as their respective channel conditions are not known a priori at the AP. Different approaches to solve this problem have been proposed. In \cite{wei2016approximate}, a message passing algorithm was proposed to jointly detect user activity and their data. This problem was also solved in \cite{wang2016dynamic} using compressive sensing. In their technique, the estimated user set in each slot depends on prior information from previous transmissions. This is valid when there exist temporal correlations between user transmissions. In \cite{nikopour2013sparse}, another user detection technique was proposed using a set of orthogonal pilot sequences which are chosen uniformly at random by the set of active users. As long as every pilot sequence is chosen by only one user, users can be accurately detected at the receiver side and their channel state information can be accurately estimated. However, even when the number of pilot sequences is very large, the probability that two or more users choose the same pilot sequence is non-zero. In this case, a collision is said to have occurred as these devices cannot be distinguished by the AP. While authors in \cite{zhang2014sparse} suggest that devices adopt some back-off scheme to resolve this collision, this implies that any collision will lead to the loss of all data including those devices that did not collide. This is very wasteful of resources.

\subsection{Contributions}
Motivated by these findings, this paper investigates the performance of an uncoordinated massive NOMA scheme where user detection and channel estimation is carried out via pilot sequences that are transmitted simultaneously with the device's data. In particular, we investigate the performance of massive NOMA with collisions, which is missing from previous works in this area. If the AP is not able to resolve the collisions, they are regarded as interference. In this case, other devices that did not collide in the given slot can still be recovered. This allows for a more practical transmission scheme along with a more rounded assessment of its performance and suitability for mMTC. All in all, the work provides the following three major contributions.
\subsubsection{Grant-Free Massive NOMA Scheme}
We consider an uplink grant-free NOMA setting where devices jointly transmit a randomly chosen pilot sequence along with their data. For this setting, there is always a non-zero probability that two or more devices choose the same pilot sequence. The receiver is only able to estimate their aggregate power. However, it is unable to distinguish the devices from one and another, and a collision is said to have occurred. In this work, we propose to treat these codewords as interfering signals at the AP. We derive the distribution of the number of collided devices and show that the aggregate interference power can be well-approximated by a PPP. Finally, we present the characteristic function of the aggregate interference power which is an essential parameter in the performance analysis of this system.
\subsubsection{Outage Probability of Massive NOMA}
For the proposed framework, we first consider the case where all the devices transmit at the same fixed code-rate. We derive the expression of the outage probability for the case of joint decoding and successive interference cancellation. The evaluation of the exact expression is shown to be daunting especially for the case of joint decoding. To overcome this problem, we propose a simplified expression and demonstrate its accuracy through simulations. Our results show that the optimal length of the pilot sequences scales linearly with the packet arrival rate. Our results also show that SIC achieves a similar performance as SJD while reducing the decoding complexity.
\subsubsection{Performance Analysis for Successive Interference Cancellation}
We then consider the case where the devices transmit using rateless codes. In this case, the rate is determined on the fly and varies from slot to slot based on the system load, received powers and interferers. The receiver stops transmissions by broadcasting a beacon when the throughput is maximized. We derive the expression for the maximum throughput for the case of joint decoding and successive interference cancellation. The evaluation of the exact expression is shown to be very complicated. Based on this, we propose a simplified expression and demonstrate its accuracy through simulations. Our results show that the maximum throughput of SJD is almost double that of SIC. However, we explain that the maximum throughput under SIC is achievable in practice whereas the existence of codebooks that can achieve the maximum throughput in the case of SJD is questionable.

The rest of this paper is organized as follows. Section II describes the system model and the considered transmission scheme for uncoordinated massive NOMA. In Section III, we show that the aggregate power of collisions can be modelled as a PPP and present its distribution. In Section IV and V, we derive the outage probability and maximum system throughput for uncoordinated NOMA under massive access for SJD and SIC, respectively. Numerical results are presented in Section VI. Practical considerations are discussed in Section VII, and conclusions are drawn in Section VIII.

In this paper, we denote by $f_X(x)$, $F_X(x)$, $\psi_X(\omega)$ and $\mu_X(n)$ the probability density function (PDF), the cumulative density function (CDF), the characteristic function (CF), and the $n^{th}$ moment of $X$, respectively. The cardinality of the set $\mathcal{X}$ is denoted by $|\mathcal{X}| = X$. All logarithms are taken to the base 2, unless otherwise indicated. $C(x) \:= \log(1+x)$ is the point-to-point Gaussian channel capacity with $x$ denoting the signal-to-noise ratio (SNR), $\Gamma(\cdot)$ is the Gamma function, and $j = \sqrt{-1}$. Finally, $\delta(x)$ is the indicator function such that $\delta(x) = 1$ if $x>0$ and is zero otherwise.
\section{System Model}
\begin{table}[!t]
  \centering
  \caption{Notation Summary}
  \begin{tabular}{|p{0.9cm}| p{7cm}|}
  \hline
    \textbf{Notation}&\textbf{Description}\\
  \hline
  $M$& Number of symbols in a time slot\\
    \hline
  $K$& Number of information bits in a packet\\
  \hline
  $q$&Number of symbols in a pilot sequence\\
  \hline
  $\mathcal{N}$& Set of transmitting devices \\
  \hline
    $\mathcal{L}$&Set of pilot sequences\\
  \hline
    $\mathcal{L}_s$&Set of singleton pilot sequences/layers/devices\\
  \hline
  $\mathcal{N}_{\ell}$&Set of devices that chose the $\ell^{th}$ pilot sequence\\
  \hline
    $\mathcal{Z}$&Set of devices in collision (interfering)\\
  \hline
  $P_T$&Maximum transmit power of a device\\
  \hline
  $P_i$&Received power of device $i$\\
  \hline
  $\hat{P}_i$&Received power of the device in $\ell^{th}$ singleton layer\\
  \hline
  $\rho_{\ell}$&SINR of layer $\ell$, $\ell \in \mathcal{L}_s$\\
  \hline
    $\hat{\rho}_{\ell}$&SINR of the $\ell^{th}$ singleton layer\\
  \hline
    $R_c$& Device code rate \\
  \hline
    $R_f$&Device effective rate\\
  \hline
  \end{tabular}\label{not}
\end{table}
\subsection{Overview}
We model the location of the devices transmitting in any time slot as a homogeneous Poisson point process (PPP) on an annular region with minimum and maximum radii $d_{\text{min}}$ and $d_{\text{max}}$ \cite{4802198,andrews2011tractable,rabbachin2011cognitive,ghosh2012heterogeneous,novlan2013analytical}, respectively. The AP is located at the center, and the average number of transmitting devices surrounding the AP per time slot is denoted by $\lambda$. Devices are considered to be static, and the channel is modelled as a block fading channel. That is, the devices' channel conditions remain constant for the duration of one packet and vary randomly and independently from one slot to the other. For reciprocal channels, each device can make use of the pilot signal sent periodically over the downlink channel by the AP to synchronize their timing to that of the AP. The impact of asynchrony is discussed in Section \ref{Sec_Prac}.

For each time slot, the AP initiates uplink transmissions through beaconing. To minimize signalling overhead, we assume that the beacon signal carries load information based on which the devices adjust their code rate. The received signal at the AP for a given time slot $t$ can be expressed as:
\begin{align}
\textbf{y}^{(t)} = \sum_{i \in \mathcal{N}^{(t)}} \textbf{x}_i^{(t)} + \textbf{w}^{(t)},
\end{align}
where $\textbf{x}_{i}^{(t)}$ is the codeword transmitted by device $i$ from the total set of transmitting devices $\mathcal{N}^{(t)}$ with a received power of $P_i^{(t)}$, and $\textbf{w}^{(t)}$ is a circular symmetric white Gaussian noise with unity variance (could include inter-cell interference). We consider Rayleigh fading and a path-loss channel model. Thus, $P_i^{(t)} = |h_i^{(t)}|^2d_i^{-\alpha}P_T$, where $h_i$, $d_i$, $\alpha$ and $P_T$ denote the small scale fading gain of the $i^{th}$ device, the distance between the $i^{th}$ device and the AP, the path-loss exponent and the transmit power, respectively. In what follows, we consider tight latency requirements where no retransmission opportunities are available, i.e., if the transmission of the packet is not successful in one slot, the packet is dropped. Thus, for the remaining part of the paper, we focus our analysis on a single time-slot and, thus, we drop the superscripts.

In this work, each codeword is concatenated with a pilot sequence of length $q$ symbols for user detection and decoding. Thus, each codeword is of length $M-q$ symbols. Assuming each device has a set of $K$ information bits to transmit, the code rate per device is defined as
\begin{equation}
R_c: = \frac{K}{M-q}\quad\text{bits/symbol}.
\end{equation}
On the other hand, the effective rate per device is defined as
\begin{equation}
R_f := \frac{K}{M}\quad \text{bits/symbol}.
\end{equation}
$R_f$ represents the ratio of the number of information bits to the total number of symbols transmitted. Thus, the effective rate takes into consideration the redundancy incurred by the pilot sequence.
\subsection{Transmission Scheme}
In each slot, each device chooses a pilot sequence of length $q$ symbols independently and uniformly at random from a set $\mathcal{L} = \{1,2,...,L\}$. We categorize the pilots sequence into three types. An idle pilot sequence is a pilot sequence which has not been chosen by any device. A singleton pilot is a pilot sequence chosen by only one device, and a collision pilot sequence is a pilot sequence chosen by two or more devices. Each pilot sequence $\ell$ is made unique to a specific code book $\mathcal{C}_{\ell}$ and, thus, acts as the device's signature. A device $i$ that has chosen the pilot sequence $\ell$ encodes its data $\textbf{b}_i$ into a codeword $\textbf{x}_i = f_{\ell}(\textbf{b}_i)$.

In practice, the set of received pilot sequences is determined by the AP via the power delay profile which is constructed by performing cross-correlations between the known pilot sequences and the received signals and averaging the absolute square values of the created channel impulse responses \cite{manabe1992superresolution}. This is based on the fact that the auto-correlation of orthogonal pilot sequences can be approximated by a delta function, and its cross-correlation with other pilot sequences yields all-zero sequences \cite{manabe1992superresolution}. Thus, it is essential that the chosen set of pilot sequences have good auto and cross-correlation properties, e.g., Zadoff-chu \cite{li2007constructive}, Golden codes \cite{belfiore2005golden}, m-sequences \cite{buravcas2002efficient}. In what follows, we assume perfect pilot sequence detection and power estimation, i.e., perfect device detection and channel estimation.

\begin{figure}
  \centering
  \includegraphics[width=\columnwidth]{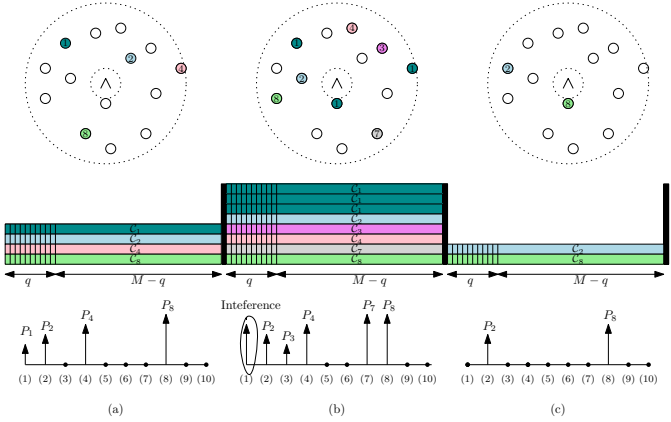}\\
  \caption{Different case scenarios of a network with $L= 10$ pilot sequences: (a) $N = 4$, (b) $N = 8$ and (c) $N = 2$.}\label{tx}
\end{figure}

Some examples are illustrated in Fig. \ref{tx}. In Fig. \ref{tx}(a), the four transmitting devices choose unique pilot sequences (1, 2, 4 and 8). Thus, there are no collisions. The received powers are estimated as $P_1$, $P_2$, $P_4$, and $P_8$. The remaining pilot sequences (3, 5, 6, 7, 9 and 10) are idle pilot sequences. In Fig. \ref{tx}(b), three devices choose the first pilot sequences. Devices that choose the same pilot sequence also choose the same code book. These devices cannot be distinguished by the AP and are said to have collided. We will refer to each code book as a layer. Thus, these devices have collided over the first layer. In what follows, we assume that the AP can distinguish between collision layers and singleton layers. However, for collision layers, the AP does not know the number of colliding devices and is, thus, unable to decode the collision layers. Instead, the AP treats them as interference whose aggregate power can be estimated from the power delay profile. The feasibility of these assumptions will be discussed in Section \ref{Sec_Prac}.

We denote by $\mathcal{N}_{\ell}$ the set of devices that have chosen the $\ell^{th}$ pilot sequence of size ${N}_{\ell}$. For a sufficiently large number of pilot sequences, the random variables ${N_1}, N_2, ...,N_{L}$ are Poisson random variables with average $\lambda/L$.  This is a practical assumption as the length of the pilot sequences need to be large enough for the AP to accurately calculate the received powers, and the number of pilot sequences increases with its length \cite{li2007constructive}. The set of devices that have collided is denoted by $\mathcal{Z} = \{i\Big| i\in\mathcal{N}_{\ell}, N_{\ell} > 1, \forall \ell\in\mathcal{L}\}$. Moreover, the set of singleton layers is denoted by $\mathcal{L}_s= \{\ell\Big|N_{\ell} = 1, \forall \ell\in\mathcal{L}\}$. Then, the equivalent received signal as seen by the AP can be expressed as
\begin{align}
\textbf{y} = \sum_{n\in\mathcal{N}_{\ell},\ell \in \mathcal{L}_s} {\textbf{x}}_{n} + \sum_{n' \in \mathcal{Z}}{\textbf{x}}_{n'} +  \textbf{w},
\end{align}
where ${\textbf{x}}_{i} = f_{\ell}(\textbf{b}_i)$ for ${i\in\mathcal{N}_{\ell}}$. The received signal-to-interference-and-noise-ratio (SINR) of the singleton layers is given as
\begin{align}\label{SINR_exp}
\rho_{\ell} = \frac{P_i}{I + \sigma^2},\quad \text{for }i\in\mathcal{N}_{\ell},\ell\in\mathcal{L}_s,
\end{align}
where $I$ denotes the aggregate interference power caused by these collisions given as:
\begin{equation}\label{I_exp}
I = \sum_{n\in \mathcal{Z}}|h_n|^2d_n^{-{\alpha}}{P_T}.
\end{equation}
It is worthy of pointing out the main differences between this transmission scheme and that of Long-term Evolution (LTE). In LTE \cite{wiriaatmadja2015hybrid}, the set of orthogonal sequences used for user detection are called preambles. Users transmit their chosen preambles separately from their data on a dedicated random access channel, and the AP allocates resources on the uplink data channel for each of the detected preambles. A collision occurs when two or more devices choose the same preamble. These devices will transmit their packets over the same time-frequency resources, and the AP will not be able to decode them correctly. Although many techniques have been proposed to resolve these collisions, e.g. \cite{zhang2016uplink}, LTE remains a form of coordinated orthogonal transmission scheme and, thus, is not suitable for massive access as explained in Section I. More importantly, the control signals and the data are transmitted over separate channels at different times which was shown to be strictly sub-optimal for small payloads \cite{durisi2016toward}.
\section{Preliminaries}
In this section, we first derive the distribution for the received power of a single user randomly located in the cell. Then, for Poisson packet arrivals, we show that the colliding/interfering devices can be well-approximated by a Poisson point process (PPP). Based on this, we find the characteristic function (CF) for the aggregate power $I$ defined as $\psi_I(j\omega) := \mathbb{E}\left[e^{-j\omega I}\right]$. These parameters will be useful in evaluating the outage probability and the throughput of the considered uncoordinated NOMA setting.
\subsection{Distribution of Received Power}
For the case where devices always transmit with maximum transmit power $P_T$ over a channel subject to Rayleigh fading and path-loss, the distribution of the received power of a singleton layer is given in the following corollary. We refer the readers to Appendix \ref{app:1} for the proof. We make note that the extension of this to Nakagami fading or any other fading distribution is straight-forward.
\begin{corollary}
For a device uniformly distributed in an annulus with a minimum radius of $d_{\min}$ and a maximum radius of $d_{\max}$, the CDF of the received power $P = |h|^2d^{-\alpha}$ ($P_T = 1$) at the origin under Rayleigh fading is
\begin{multline}\label{CDF_P_NoCSIT}
F_P(p) = 1 - \frac{p^{-\frac{2}{\alpha}}\Gamma\left[\frac{2}{\alpha}+ 1\right]}{d_{\max}^2 - d_{\min}^2} + \frac{d_{\min}^2}{d_{\max}^2 - d_{\min}^2}.
\end{multline}
\end{corollary}
\subsection{Aggregate Interference Power}
\begin{figure}
  \centering
  \includegraphics[width=\columnwidth]{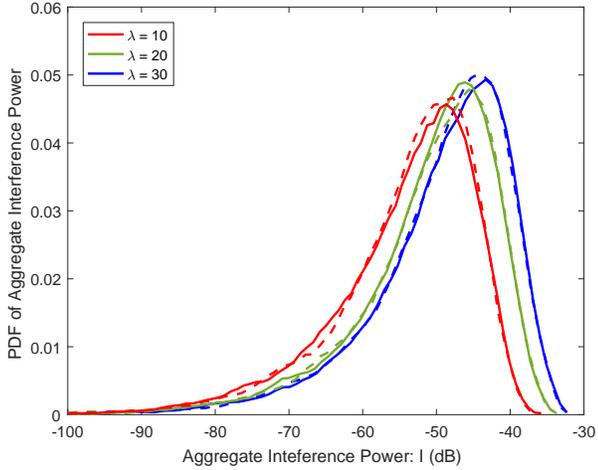}\\
  \caption{Aggregate interference power for $L-L_s = 200$. The solid lines correspond to the actual distribution, and the dashed lines correspond to the Poisson distribution. Readers are referred to Table II for the remaining system parameters.}\label{proof_PPP}
\end{figure}

We now proceed to find the distribution of the aggregate interference power $I$. For that, we first find the exact distribution of the number of interferers in the following lemma, i.e., number of devices in collision $Z$. We refer the readers to Appendix \ref{app:2} for the proof.
\begin{lemma}\label{lemma_prob_n_colliding}
Consider a total of $L$ layers where packet arrivals over each layer are Poisson distributed with an average of $\lambda/L$. Given that the number of singleton layers is $L_s<L$, the probability of having a total of $n$ packets collide in a given time slot is expressed in (\ref{prob_n_colliding}) (top of the next page).
\begin{figure*}
\begin{equation}\label{prob_n_colliding}
\text{Pr}\left(Z = n\Big| L_s\right) =  \frac{\left(\frac{(L-L_s) \lambda}{L}\right)^{n}e^{-\frac{(L-L_s) \lambda}{L}}}{n!(1-\lambda e^{-\lambda})^{L-L_s}}\left(1 + \sum_{c = 1}^{\min\{L-L_s-1,n\}}\begin{pmatrix} L-L_s\\c \end{pmatrix}\left(-\lambda e^{-\lambda}\right)^c \frac{\left((L-L_s-c)\frac{\lambda}{L}\right)^{n-c} e^{-(L-L_s-c)\frac{\lambda}{L}}}{(n -c)!}\right).
\end{equation}
\hrulefill
\end{figure*}
\end{lemma}
Although $Z$ devices are only Poisson distributed in space and not in number, our results in Fig. \ref{proof_PPP} show that their aggregate power can be well-approximated by a PPP. Accordingly, the distribution of the aggregate interference power $I$, given $L_s$, can be well-approximated by the skewed truncated stable distribution \cite{wildemeersch2013cognitive}, and its CF can be expressed as
\begin{equation}
\psi_I(j\omega) = e^{\gamma_I\Gamma(-a_I)\left({(g_I-j\omega)^{\alpha_I}}  - g_I^{\alpha_I}\right)},
\end{equation}
where the parameters $\alpha_I$, $g_I$ and $\gamma_I$ determine the shape of the distribution. In particular, the parameters $\alpha_I$ and $\gamma_I$ are related to the dispersion and the characteristic exponent of the stable distribution, respectively, and the parameter $g_I$ is the argument of the exponential function used to smooth the tail of the stable distribution. These parameters are found through the method of the cumulants \cite{wildemeersch2013cognitive} and are given as
\begin{align}
&\alpha_{I} = \frac{2}{\alpha},\\
&g_I = \frac{\kappa_I(1)(1-\alpha_I)}{\kappa_I(2)},\\
&\gamma_I = \frac{-\kappa_I(1)}{\Gamma[-\alpha_I]\alpha_I\left(\frac{\kappa_I(1)(1-\alpha_I)}{\kappa_I(2)}\right)^{\alpha_I-1}}.
\end{align}
Here, $\kappa_I(n)$ denotes the $n^{th}$ cumulant of the interference power and is given as
\begin{align}
\kappa_{I}(n) = \frac{2 \lambda_Z}{n\alpha - 2}\frac{d_{\min}^{2-n\alpha} - d_{\max}^{2-n\alpha}}{d_{\text{max}}^2-d_{\text{min}}^2}\mu_v(n)P_T^{n},
\end{align}
where $v = |h|^2$ is a chi-squared distributed random variable under Rayleigh-fading, $\mu_v(n)$ is its $n^{th}$ moment, and $\lambda_Z(L_s):= \mathbb{E}[Z\big|L_s]$.

Finally, the inversion theorem \cite{wildemeersch2013cognitive} dictates that the cumulative-density function (CDF) of $I$ can be computed as
\begin{equation}\label{inv_theorem}
F_{I}(x) = \frac{1}{2} - \frac{1}{2\pi}\int_0^{\infty} \text{Re}\left\{ \frac{\psi_I(-j\omega)e^{j\omega x}- \psi_I(j\omega)e^{-j\omega x}}{j\omega}\right\}\text{d}\omega,
\end{equation}
and the probability density function (PDF) of $I$ can be computed as
\begin{equation}
f_I(x) = \frac{1}{2\pi}\int_0^{\infty}e^{-j\omega x}\psi_I(-j\omega)d\omega.
\end{equation}

\section{Performance of Massive NOMA with Successive Joint Decoding}\label{Sec SJ}
In this section, we consider the case where all the singleton layers are jointly decoded at the receiver using a maximum likelihood decoder. Although jointly decoding all of the singleton layers is optimal in terms of the common outage event, i.e., the probability that at least one layer decoded incorrectly, it is not optimal in terms of the individual outage event, i.e., the probability that one layer is decoded incorrectly. Therefore, we consider the successive joint decoder (SJD) where a subset of the layers can be decoded jointly while regarding the remaining layers as interference. For that, we derive the expressions for the outage probability and the throughput. For ease of notation, we define $\hat{\textbf{P}} := [\hat{P}_{\ell}]_{1\leq \ell \leq L_s}$, where $\hat{P}_{\ell}$ is the received power of the device in the $\ell^{th}$ singleton layer. Similarly, we define $\hat{\boldsymbol{\rho}} := [\hat{\rho}_{\ell}]_{1\leq \ell \leq L_s}$, where $\hat{\rho}_{\ell}$ is the SINR of the $\ell^{th}$ singleton layer.
\subsection{Outage Probability with SJD}\label{Subsec_SJD_Outage}
Consider the case where all the devices encode their data using a fixed rate code with code rate $R_c := \frac{K}{M-q}$. In this case, we are interested in characterizing and evaluating the outage probability of the system. That is, the probability that a device is not successfully decoded in a given time slot. Given $L_s$ singleton layers with a received power vector $\hat{\textbf{P}}$, subject to an interference power $I$, a successive joint decoder first tries to decode all $L_s$ layers jointly. However, if the equal-rate point $(R,...,R)\in\mathbb{R}^{L_s}_+$ is outside the $L_s$-dimensional capacity region defined by the received SINRs $\hat{\boldsymbol{\rho}}$, the AP will try  to decode the strongest $L_s-1$ devices by treating the weakest layer as interference. The weakest layer is defined as the singleton layer with the smallest received power. Similarly, if the equal-rate point $(R,...,R)\in\mathbb{R}^{L_s-1}_+$ is outside the ${L}_s-1$-dimensional capacity region, the AP will try  to decode the strongest ${L}_s-2$ layers by treating the two weakest layers as interference. The process repeats until decoding is successful or until there are no more layers to decode. Now, consider the descending ordered set $\hat{P}_{(1)}\geq \hat{P}_{(2)}\geq \ldots \geq \hat{P}_{(L_s)}$.

For that, the outage probability is defined as the average number of singleton layers that cannot be decoded successfully. Given $L_s$ singleton layers with a received power vector $\textbf{P}$, subject to an interference power $I$, the fraction of layers in outage can be expressed as
\begin{align*}
&\epsilon_{\text{SJD}}(L_s,I,\hat{\textbf{P}}) = 1 - \max_{0\leq \ell\leq L_s} \frac{\ell}{L_s}, \numberthis \label{1_e_SD_asym}\\
&\text{s.t. } R_c \leq  \min_{1\leq i \leq \ell}\frac{1}{i}\log \left(1+\frac{\sum_{c=\ell-i+1}^{\ell}\hat{P}_{(c)}}{\sum_{c'=\ell + 1}^{L_s} \hat{P}_{(c')} + I + \sigma^2}\right).
\end{align*}
Here, $\ell$ denotes the maximum number of layers that can be decoded out of the $L_s$ singleton layers. In general, the outage probability can be expressed as
\begin{align}\label{raw_e_SJD}
\epsilon_{\text{SJD}} = 1 - \mathbb{E}_{L_s,I,\hat{\textbf{P}}}\left[\sum_{\ell= 0}^{L_s} \frac{\ell}{L_s}\text{Pr}\left(\bar{\phi}_{L_s,L_s},\ldots,\bar{\phi}_{\ell+1,L_s},{\phi}_{\ell,L_s}\right)\right],
\end{align}
where $\phi_{\ell,L_s} =$
\begin{align*}
\delta\left( R_c \leq \min_{1\leq i\leq \ell}\frac{1}{i} \log \left(1+\frac{\sum_{c=\ell-i+1}^{\ell}\hat{P}_{(c)}}{\sum_{c'=\ell + 1}^{L_s} \hat{P}_{(c')}+I + \sigma^2}\right)\right),
\end{align*}
Thus, to find the outage probability, not only do we need to average over the different number of singleton layers as well as the different aggregate interference powers, we also need to average over the numerous realizations of the received power vector. Thus, computing the outage probability for uncoordinated multiple access is even more daunting than that for coordinated multiple access \cite{tuninetti2002throughput}. In what follows, we make use of the massive aspect of mMTC to simplify this expression.

Consider a set of $L$ i.i.d. random variables $\left[X_i\right]_{1\leq i\leq L}$, where $X_1 \geq X_2 \geq ... \geq X_L$. Then, from ordered statistics \cite{orderedStatistics}, we have
\begin{align}\label{ordered_theorem}
\lim_{L\rightarrow \infty} \text{Pr}\left(|X_{\lfloor yL\rfloor} - F^{-1}_X(1-y)|> \nu \right) = 0,
\end{align}
for all $y\in[0,1]$ and $\nu > 0$. Based on this, for an asymptotically large number of singleton layers, the following equality holds for all $\ell_1,\ell_2 \in[1,L_s]$.
\begin{align*}
\lim_{L_s\rightarrow \infty} \sum_{i = \ell_1}^{\ell_2}  \hat{P}_{(i)} =&
 L_s\int_{\ell_1/L_s}^{\ell_2/L_s} F^{-1}_P(1-y)\text{d}y  \\  =& L_s\left( g(\ell_2/L_s)  - g(\ell_1/L_s)\right),\numberthis \label{prop_ordered}
\end{align*}
where $g(y) := \int F^{-1}_P(1-y)\text{d}y$. For Rayleigh fading and a path-loss channel model, $g(y)$ is evaluated by integrating the inverse of (\ref{CDF_P_NoCSIT}) and is expressed in (\ref{g_y}).
\begin{figure*}
\begin{equation}\label{g_y}
g(y) = \frac{2\left((d_{\max}^2-d_{\min}^2)y - d_{\min}^2\right)}{(\alpha-2)(d_{\max}^2-d_{\min}^2)\left(\Gamma\left[\frac{2}{\alpha}+1\right]\right)^{\frac{-\alpha}{2}}\left((d_{\max}^2-d_{\min}^2)y - d_{\min}^2\right)^{\frac{\alpha}{2}}}.
\end{equation}
\hrulefill
\end{figure*}
For the special case of Rayleigh fading only, we have
\begin{align}
g_{\text{Ray.}}(y) = (1-y)\log_e(1-y) - (1-y).
\end{align}
Using this property, the complexity of computing the outage probability of massive NOMA in (\ref{raw_e_SJD}) is significantly reduced as $\textbf{P}$ converges to a deterministic value as $L_s \rightarrow \infty$. The expression for that is derived in the following lemma. We refer the readers to Appendix \ref{app:3} for the proof. We make note that in practice, we find that this property gives sufficiently accurate results for $L_s > 10$.
\begin{lemma}
Consider a time slot of length $M$ symbol durations. For a code rate $R_c$, $L$ pilot sequences of length $q$ and a packet arrival rate of $\lambda$ packets, the outage probability of the massive NOMA system with SJD can be expressed as
\begin{multline}\label{fig_e_SD_asym}
\epsilon_{\text{SJD}}= 1 - \mathbb{E}_{L_s}\left[\int_0^1\frac{v}{2\pi}\int_0^{\infty}e^{-j\omega I^*_{\text{SJD}}(v)}\psi_I(-j\omega)d\omega\right],
\end{multline}
where $I_{\text{SJD}}^*(v)$ is the solution to the equation below
\begin{multline}\label{I_SD}
L_sR_c - \min_{0< u \leq 1}\frac{1}{uv} \log \left(1+\frac{g(v) - g(v-uv)}{g(1)-g(v)+\frac{I + \sigma^2}{L_s}}\right) = 0.
\end{multline}
\end{lemma}
\noindent Here, $I^*_{\text{SJD}}(v)$ denotes the largest interference power for which the largest fraction of layers that can be decoded successfully is $v$. The integral in Lemma 3 averages over the aggregate interference power, and the expectation averages over the number of singleton layers.
\subsection{Maximum Throughput with SJD}\label{SJD-zeta}
We now consider the case where devices can transmit as many coded symbols as necessary. This is feasible when devices use rateless channel codes to encode their information \cite{shirvanimoghaddam2017fundamental}. With rateless codes, the length of the codeword is determined on the fly and is adaptive to the load, received powers and interference power. Thus, the duration of the slot ($M$) as well as the devices' code rate ($R_c$) also varies from one slot to the other. Transmissions are stopped by the AP when the throughput is maximized through beaconing. The throughput is defined as the ratio of the number of successfully decoded information bits to the length of the transmitted codeword ($M-q$). This is determined by the AP and can be made known to the devices through beacons. Given $L_s$ singleton layers, a received power vector $\hat{\textbf{P}}$, subject to an interference power $I$, the maximum instantaneous throughput is given as
\begin{multline*}
\zeta_{\text{SJD}}(L_s,I,\hat{\textbf{P}}) = \\\max_{1\leq \ell\leq L_s} \min_{1\leq i\leq \ell} \frac{\ell}{i}\log\left(1 + \frac{\sum_{c=\ell-i+1}^{\ell}\hat{P}_{(c)}}{\sum_{c'=\ell + 1}^{L_s} \hat{P}_{(c')}+I + \sigma^2}\right).
\end{multline*}
As the throughput varies from slot to slot, the maximum average throughput should be averaged over all possible values of $L_s$, $I$ and $\hat{\textbf{P}}$. However, from (\ref{ordered_theorem}), the maximum instantaneous system throughput under massive access can be expressed from \cite{tuninetti2002throughput} as
\begin{multline}
\zeta_{\text{SJD}}(L_s,I,\hat{\textbf{P}}) = \\\max_{0\leq v \leq 1}\min_{0<u\leq 1} \frac{1}{u}\log \left(1+\frac{g(v) - g(v-uv)}{g(1) - g(v) + \frac{I + \sigma^2}{L_s}}\right).
\end{multline}
Then, the average maximum system throughput under massive access is derived in the following lemma. We refer the readers to Appendix \ref{app:4} for the proof.
\begin{lemma}
For $L$ pilot sequences of length $q$ and  a packet arrival rate of $\lambda$ packets, the average maximum system throughput of massive NOMA with SJD can be expressed as in (\ref{1_zeta_SD_asymp}).
\begin{figure*}
\begin{align}\label{1_zeta_SD_asymp}
\zeta_{\text{SJD}}=
\mathbb{E}_{L_s}\left[\max_{0\leq v \leq 1}\min_{0<u\leq 1}\frac{1}{u}\int_0^{\infty}\frac{\psi_I(-s)}{s}\left(e^{-s((g(1) - g(v))L_s +\sigma^2)}-e^{-s ((g(1) - g(v-uv))L_s +\sigma^2}\right)\text{d}s \right].
\end{align}
\hrulefill
\end{figure*}
\end{lemma}
\section{Performance of Massive NOMA with Successive Interference Cancellation}
As joint decoding is often infeasible in practice due to complexity constraints, we consider successive interference cancellation (SIC) in this section. In each stage of SIC, the strongest layer is decoded by regarding remaining layers as interference. Thus, we only need a single-user decoder. If the layer is decoded successfully, the decoded layer is subtracted from the received signal. We assume perfect SIC. In the second stage, the second strongest layer is decoded by regarding the remaining layers as interference. SIC stops when a layer is decoded unsuccessfully or when there are no more layers to decode.
\subsection{Outage Probability with SIC}
Given that the strongest $\ell - 1$ layers have been decoded successfully, the $\ell^{th}$ strongest layer is decoded successfully if the following condition is true.
\begin{equation}
\frac{\hat{P}_{(\ell)}}{\sum_{c = \ell + 1}^{L_s}\hat{P}_{(c)} + I + \sigma^2} \geq 2^{R_c} -1
\end{equation}
Thus, given that the strongest $\ell - 1$ layers have been decoded successfully, the probability that the $\ell^{th}$ strongest layer is decoded successfully is given as
\begin{align*}
&\text{Pr}\left(I\leq\frac{\hat{P}_{(\ell)} - \left(2^{R_c}-1\right)\left(\sum_{c = \ell + 1}^{L_s}\hat{P}_{(c)} +  \sigma^2\right)}{2^{R_c}-1}\right)
\end{align*}
From (\ref{ordered_theorem}), given $L_s$ and $I$, the SINR at each stage of the SIC is deterministic under massive access. Based on this, the outage probability under SIC is derived in the following proposition. The definition of the outage probability is as given in Section \ref{Subsec_SJD_Outage}, and the proof of the proposition is similar to that of Lemma 2.
\begin{proposition}
For a code rate $R_c$, $L$ pilot sequences of length $q$ and  a packet arrival rate of $\lambda$ packets, the outage probability of massive NOMA with SIC can be expressed as
\begin{multline}\label{fig_e_SIC_asym}
\epsilon_{\text{SIC}}(R) = 1 - \mathbb{E}_{L_s}\left[\int_0^1\frac{v}{2\pi}\int_0^{\infty}e^{-j\omega I^*_{\text{SIC}}(v)}\psi_I(-j\omega)d\omega\right],
\end{multline}
where
\begin{align}
I^*_{\text{SIC}}(v) := \frac{F^{-1}_P(v) - \left(2^{R_c}-1\right)\left((g(1) - g(v))L_s +  \sigma^2\right)}{2^{R_c}-1}.
\end{align}
\end{proposition}
\noindent Here, $I^*_{\text{SIC}}(v)$ denotes the largest interference power for which the largest fraction of layers that can be decoded successfully is $v$.
\subsection{Maximum Throughput with SIC}
We now consider the same setup in Section \ref{SJD-zeta}. Given $L_s$ singleton layers with a received power vector $\hat{\textbf{P}}$, subject to an interference power $I$, the maximum instantaneous throughput is given as
\begin{multline}
\zeta_{\text{SIC}}\left(\textbf{P},L_s,I\right) =\\ Ls\max_{0 < v \leq 1} v  \log\left(1 + \min_{0 < v' \leq v}\frac{F^{-1}_P(1-v')}{(g(1)-g(v'))L_s + I + \sigma^2} \right).
\end{multline}
\begin{figure*}
\begin{align}\label{1_zeta_SIC_asymp}
\zeta_{\text{SIC}} = \mathbb{E}_{L_s}\left[\max_{0 < v \leq 1} vL_s  \min_{0 < v' \leq v}\int_0^{\infty}\frac{\psi_I(-s)}{s}{e^{-s((g(1) - g(v'))L_s +\sigma^2)}}\left(1-e^{-s F^{-1}_P(1-v')}\right)\text{d}s\right].
\end{align}
\hrulefill
\end{figure*}
The average maximum system throughput is given in the following proposition. The proof of this proposition is similar to that of Lemma 3.
\begin{proposition}
For $L$ pilot sequences of length $q$ and with a packet arrival rate of $\lambda$ packets, the average maximum system throughput of massive NOMA under SIC can be evaluated from (\ref{1_zeta_SIC_asymp}).
\end{proposition}
\section{Numerical Results}
\begin{table}[!t]
  \centering
  \caption{System Parameters}
  \begin{tabular}{|p{4cm}| p{2cm} |}
  \hline
    \textbf{Parameter}&\textbf{Value}\\
  \hline
  Cell Radius&500 m\\
  \hline
  Reference Distance&50 m\\
  \hline
  Bandwidth&15 kHz\\
  \hline
  Transmit Power&23 dBm\\
  \hline
  Thermal Noise Density& -174 dBm/Hz\\
  \hline
 Receiver Noise Figure&3 dB\\
  \hline
  Path loss exponent&3.5\\
  \hline
  \end{tabular}\label{sim_par}
\end{table}
In our simulations, we consider single-tone transmissions with the system parameters listed in Table II \cite{adhikary2016performance,wang2017primer,oh2017efficient}. The number of pilot sequences $L$ is determined such that $1 - e^{-\frac{\lambda}{L}} = \beta$, where $\beta$ denotes the probability that a device suffers from collision. We also assume that $q = L$ \cite{li2007constructive,belfiore2005golden,buravcas2002efficient}. This means that the overhead associated with channel estimation and user detection scales linearly with the packet arrival rate. In what follows, we evaluate the performance of massive NOMA for different collision probabilities.

\begin{figure*}
  \centering
  \includegraphics[width=\textwidth]{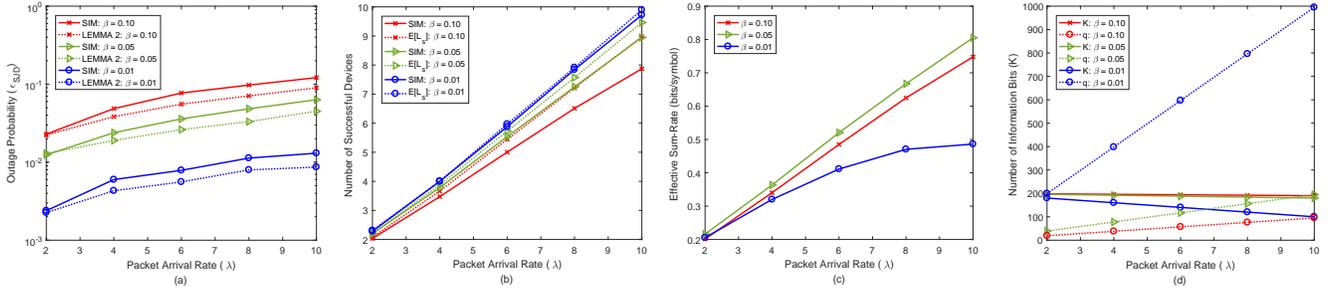}\\
  \caption{Outage Probability for massive NOMA with SJD for different values of $\beta$. The code rate $\frac{K}{M-q}$ is equal to 0.1 and the slot duration is $M = 2000$ symbols.}\label{plot_SJDe}
\end{figure*}
\begin{figure*}
  \centering
  \includegraphics[width=\textwidth]{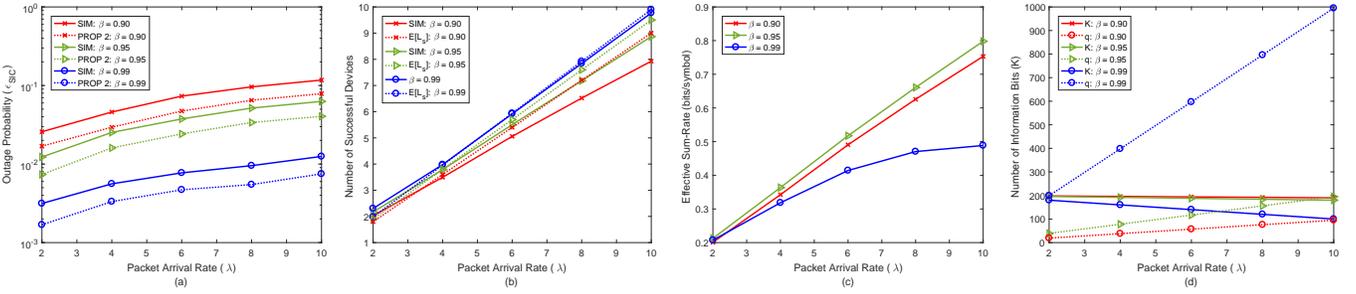}\\
  \caption{Outage Probability for massive NOMA with SIC for different values of $\beta$. The code rate $\frac{K}{M-q}$ is equal to 0.1 and the slot duration is $M = 2000$ symbols.}\label{plot_SICe}
\end{figure*}
First, we evaluate the outage probability for massive NOMA. Each time slot is of length $M = 2000$ \cite{adhikary2016performance} over which the devices transmit with a code rate of $0.1$. Results for SJD and SIC are shown in Fig. \ref{plot_SJDe} and Fig. \ref{plot_SICe}, respectively. We first observe that the analytical results from Lemma 3 and Proposition 1(PROP 1) provide a good approximation of the actual performance obtained from Monte Carlo simulations. When comparing the performance for different collision probabilities, we find that the gap between the number of successfully decoded devices and the number of singleton layers becomes more significant as $\beta$ increases for both SJD and SIC. This is because collisions in NOMA act as interference to the remaining singleton layers which leads to more outages.

In Fig. \ref{plot_SJDe}(c) and Fig. \ref{plot_SICe}(c), the effective sum-rate is evaluated as the product of the number of successfully decoded devices and the effective rate $\frac{K}{M}$. In general, if the number of pilot sequences is too small, the number of singleton layers will be low. Thus, the number of successfully decoded devices will be low as well. On the other hand, when the number/length of pilot sequences is too large, the overhead will be large. Thus, the effective rate will be low. Interestingly, it seems that the optimal value of $L$ scales linearly with the packet arrival rate. Finally, in Fig. \ref{plot_SJDe}(d) and Fig. \ref{plot_SICe}(d) we plot the maximum payload sizes that can be transmitted and compare them to the overhead induced by the pilot sequences. In general, we can say that for applications with payloads of size 100-200 bits and a target outage probability of $\approx 0.01$, we can simultaneously support up to 7 devices with SJD and up to 6 devices with SIC.

\begin{figure*}
  \centering
  \includegraphics[width=\textwidth]{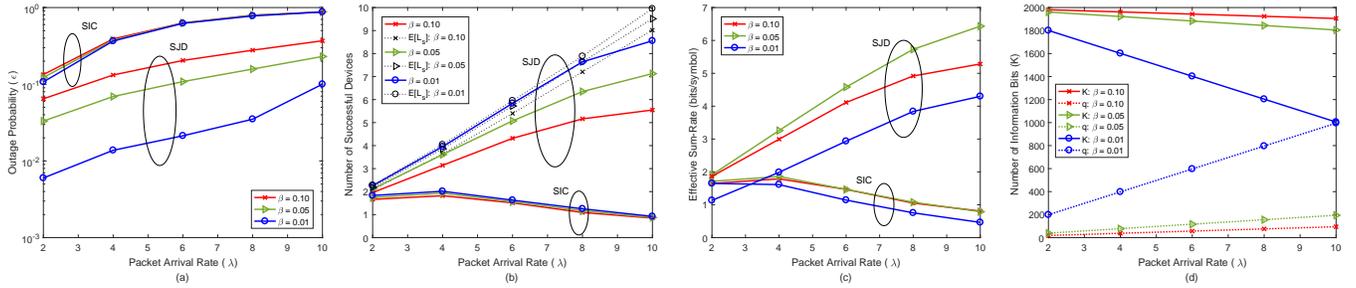}\\
  \caption{Outage Probability for massive NOMA with SJD and SIC for different values of $\beta$. The code rate $\frac{K}{M-q}$ is equal to 1 and the slot duration is $M = 2000$ symbols.}\label{plot_e2}
\end{figure*}
In Fig. \ref{plot_e2}, we evaluate the outage probability for a larger code rate ($R_c = 1$). Here, we notice that the performance gap between SJD and SIC is more significant. In fact, from Fig. \ref{plot_e2}(b), we can see that SIC can barely decode any packet. However, although the outage probability of SJD is better, it is still too high for any practical use. Moreover, as mMTC are characterized by their low data rate transmissions, SIC seems to be a good candidate for our uncoordinated massive NOMA scheme when low processing complexity is needed.

\begin{figure*}
  \centering
  \includegraphics[width=\textwidth]{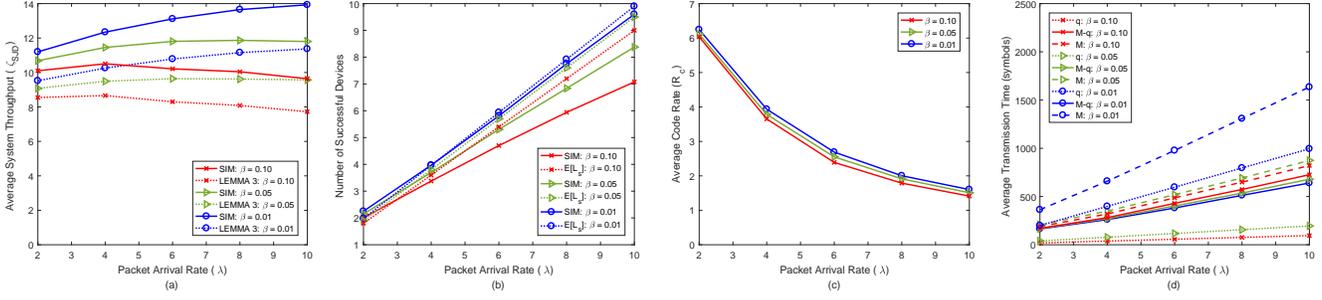}\\
  \caption{Average system Throughput for massive NOMA with SJD for different values of $\beta$ ($K = 1024$).}\label{plot_SJDr}
\end{figure*}
\begin{figure*}
  \centering
  \includegraphics[width=\textwidth]{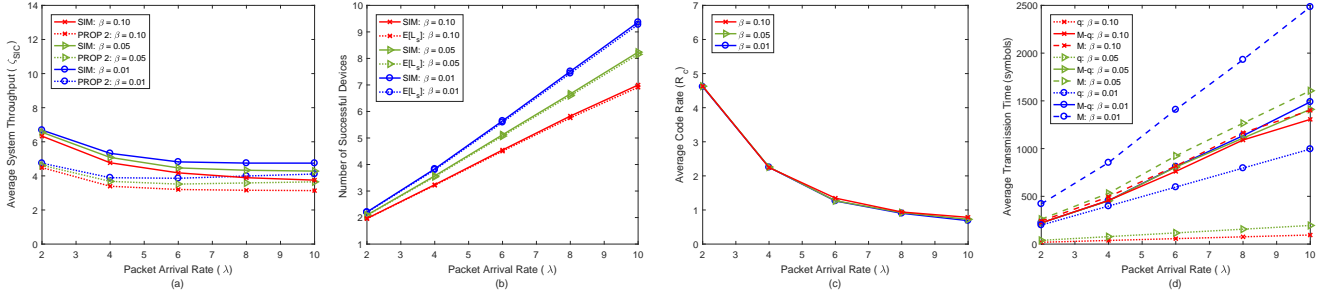}\\
  \caption{Average system Throughput for massive NOMA with SIC for different values of $\beta$ ($K = 1024$).}\label{plot_SICr}
\end{figure*}
Next, we evaluate the throughput for massive NOMA in Fig. \ref{plot_SJDr} and Fig. \ref{plot_SICr}. In this setup, each device has a set of 1024 information bits to transmit \cite{adhikary2016performance}. The number and length of pilot sequences is determined as before. We observe that the analytical results from Lemma 4 and Proposition 2 (PROP 2) also provide a good approximation of the actual performance obtained from Monte Carlo simulations. We also observe that the throughput gain of SJD is almost double of that of SIC, thus, SJD is in theory far more superior in this case. However, in the following section, we explain that
the maximum throughput under SIC is achievable in practice whereas the achievability in the case of SJD is questionable. Finally, we make note that in practice $M$ should not be larger than the coherence time. In this case, collision probabilities lower than 0.01 cannot be supported by this scheme as it will require longer pilot sequences.
\section{Practical Considerations}\label{Sec_Prac}
\subsection{Detection of Collision Layers and Achievability}
For SIC, the detection of collision layers can be implemented through a simple algorithm. At first, the AP assumes that all the layers are singleton layers and orders the layers according to their respective received powers in descending order. In the first iteration, the AP attempts to decode the strongest layer. The AP can verify that the decoded information is correct through a simple CRC-check. If decoding is successful, the decoded signal is cancelled from the received signal. If not, the AP assumes that this is a collision layer. In the second iteration, the AP attempts to decode the second strongest layer while regarding the previously undecoded layer as interference. The AP can continue to follow the same steps until there are no more layers to decode. Detecting collision layers for SJD is also possible at the expense of a larger complexity, as the AP would need decode all possible subsets of layers while regarding the remaining layers are interference. When decoding finishes, the AP would choose the largest subset of successfully decoded layers.

It is straight-forward to see that the derived bounds for SIC are achievable when capacity achieving point to point channel codes are used. However, for SJD, it is uncertain whether their exists channel codes suitable for grant-free coordinated access with joint decoding. That is mainly because these codes would not only have to adapt to the different channel conditions but also to the different channel loads. Finally, we would like to point out that there exists practical low-complex decoders such as belief-propagation-based decoders that can score performances close to that of maximum likelihood decoding
\subsection{Synchronization}
In conventional communication systems, synchronization takes place beforehand. However, it has been well established now that for massive access control signals should be minimal if not null. We would like to point out that as most MTDs are stationary devices, their timing advance need not be communicated very frequently. Alternatively, time asynchrony can provide another contention unit by which devices can be distinguished from one another without having to increase the pilot sequence length and thus without introducing redundancy. Moreover, time asynchrony can help in detecting the number of devices in collision layers. This allows the opporutnity to resolve collisions layers and decode the packets of the involved devices. However, the performance analysis for this case scenario is not straightforward as the maximum achievable rate with code-reuse is strictly lower than that with unique codes and is dependent on the codes used \cite{yang2017non}.

In general, the AP can feedback a timing advance signal in the acknowledgement. For that, the operator can either choose to operate either in a fully synchronous mode or allow the devices to introduce random asynchrony into their transmissions to create more contention units \cite{async_mac}.
\section{Conclusion}
In this paper, we considered a massive uncoordinated NOMA setting where devices choose pilot sequences from a predetermined set as their signature. Then, each device encodes its data using the pilot as the signature and transmits its selected pilot and data simultaneously with the rest of the devices. In our proposed scheme, a collision occurs when more than one device choose the same pilot sequence. The set of collided packets are treated as interference. For that, we show that the aggregate interference power can be well-approximated by a PPP. Based on this, we derive the outage probability and the maximum system throughput under successive joint decoding (SJD) as well as successive interference cancellation (SIC) for a Rayleigh fading and path loss channel model. We verified the accuracy of our derived expressions via simulations. Our results show that SIC performs close to that of SJD in terms of outage probability for packet arrival rates up to 10 packets per slot. In terms of throughput, although SJD scores almost double the throughput gain of that of SIC, we explained that this throughput might not be achievable with practical modulation and channel coding schemes.
\appendices
\section{Proof of Corollary 1}\label{app:1}
For a device uniformly distributed in an annulus of minimum radius $d_{\min}$ and maximum radius $d_{\max}$, the CDF of the distance $d$ with respect to the origin follows the truncated Pareto distribution below
\begin{align*}
F_D(d) = \frac{d^2-d_{\min}^2}{d_{\max}^2 - d_{\min}^2}.
\end{align*}
For $X = d^{-\alpha}$,
\begin{align*}
F_X(x) &= \text{Pr}(d^{-\alpha}<x)\\
&=\text{Pr}(d> x^{-\frac{1}{\alpha}})\\
&= 1 - \frac{x^{-\frac{2}{\alpha}}-d_{\min}^2}{d_{\max}^2 - d_{\min}^2}
\end{align*}
For Rayleigh fading, we can write
\begin{align*}
F_{P}(p)  &= \int_0^{\infty}\text{Pr}(vd^{-\alpha}< p|v)\text{Pr}(v)dv\\
&= \int_{0}^{\infty}\left(1 - \frac{\left(\frac{p}{v}\right)^{-\frac{2}{\alpha}} - d_{\min}^2}{d_{\max}^2 - d_{\min}^2}\right)e^{-v}\text{d}v.\\
&= 1 - \int_{0}^{\infty} \frac{\left(\frac{p}{v}\right)^{-\frac{2}{\alpha}} - d_{\min}^2}{d_{\max}^2 - d_{\min}^2}e^{-v}\text{d}v.\\
&= 1 - \frac{p^{-\frac{2}{\alpha}}}{d_{\max}^2 - d_{\min}^2}\Gamma\left[\frac{2}{\alpha}+1\right] + \frac{d_{\min}^2}{d_{\max}^2 - d_{\min}^2}.\\
\end{align*}
\section{Proof of Lemma 2}\label{app:2}
Consider a vector $\textbf{N} = \{N_i\}_{1\leq i\leq L}$, where $N_i \in \mathbb{N}$ are independent and identically distributed Poisson distributed random variables with average $\lambda$. For that, we define the following events
\begin{align*}
&A_i:&&\text{Event of }N_i = 1.\\
&B:  &&\text{Event of }N_i\neq 1, \forall i.\\
&S_n:&&\text{Event of }\sum_{i=1}^L N_i= n.
\end{align*}
Following from the definition of Poisson distributions, we have
\begin{align*}
&\text{Pr}(A_i) = \text{Pr}(A) = \lambda e^{-\lambda},\quad \forall i,\\
&\text{Pr}(B) =  \left(1-\text{Pr}(A)\right)^L = \left(1-\lambda e^{-\lambda}\right)^L,\text{ and}\\
&\text{Pr}\left(S_n\right) = \frac{\left(L\lambda\right)^ne^{-L\lambda}}{n!}.
\end{align*}
The last equation follows from the fact that the sum of any $c$ elements of $\textbf{N}$ is a Poisson distributed random variable with average $c\lambda$.

We want to calculate the probability that the sum of these elements is equal to some positive integer $n$ given that $N_i\neq 1$ for $1\leq i\leq L$. This can be expressed as
\begin{align*}
\text{Pr}\left(S_n |B\right) = \frac{\text{Pr}\left(B| S_n\right)\text{Pr}\left(S_n\right)}{\text{Pr}\left(B\right)} = \frac{\left(1-\text{Pr}\left(\bar{B}| S_n\right)\right)\text{Pr}\left(S_n\right)}{1 - \text{Pr}\left(\bar{B}\right)}.
\end{align*}
Moreover, we have
\begin{multline*}
\text{Pr}\left(A_1\cap A_2 \cap ... A_{c}\Big|S_n\right) = \text{Pr}\left(A\right)^{c}\frac{\left((L-c)\lambda\right)^{n-c}e^{-(L-c)\lambda}}{(n - c)!},
\end{multline*}
for $c \leq \min\{L-1,n\}$; it is zero otherwise. Given $S_n$,  the probability that at least one of the elements of vector $\textbf{N}$ is equal to 1 can be expressed as:
\begin{align*}
\text{Pr}\left(\bar{B}|S_n\right) &=\text{Pr}\left(\bigcup_{i=1}^{L} A_i\Big| S_n\right) \\
&= \sum_{c=1}^{L} (-1)^{c+1}\sum_{\substack{i_1,\ldots,i_c:\\  1\leq i_1< \ldots < i_c \leq L}} \text{Pr}\left(A_{i_1}\cap \ldots \cap A_{i_c}|S_n\right)\\
&=L\text{Pr}(A)\frac{\left((L-1)\lambda\right)^{n-1}e^{-(L-1)\lambda}}{(n - 1)!}-\\&\begin{pmatrix}L\\2\end{pmatrix}\text{Pr}\left(A\right)^2\frac{\left((L-2)\lambda\right)^{n-2}e^{-(L-2)\lambda}}{(n - 2)!} +\ldots+ \\ &(-1)^{\theta}\begin{pmatrix}L\\\theta\end{pmatrix}\text{Pr}\left(A\right)^{\theta}\frac{\left((L-\theta)\lambda\right)^{n-\theta}e^{-(L-\theta)\lambda}}{(n - \theta)!}\\
&=
\sum_{c = 1}^{\theta}\left(-1\right)^{c+1}\begin{pmatrix} L\\c \end{pmatrix} \left(\lambda e^{-\lambda}\right)^c\frac{\left((L-c)\lambda\right)^{n-c} e^{-(L-c)\lambda}}{(n -c)!},
\end{align*}
 where $\theta = \min\{L-1,n\}$. Then, it is straight-forward to arrive at Lemma \ref{lemma_prob_n_colliding} by calculating $\frac{\left(1-\text{Pr}\left(\bar{B}| S_n\right)\right)\text{Pr}\left(S_n\right)}{1 - \text{Pr}\left(\bar{B}\right)}$ and substituting $L$ by $L-L_s$ and $\lambda$ by $\frac{\lambda}{L}$.
\section{Proof of Lemma 3}\label{app:3}
\begin{figure*}
\begin{align*}
\epsilon_{\text{SJD}}\stackrel{(a)}{=}&1 - \mathbb{E}_{L_s}\left[\int_0^{\infty}\max_{v}\left\{v,R\leq \min_{0\leq u\leq 1}\frac{1}{uvL_s}\log\left( 1 + \frac{g(v) - g(v-uv)}{g(1) - g(v) + \frac{x + \sigma^2}{L_s}}\right)\right\}f_I(x)\text{d}I\right]\\
\stackrel{(b)}{=}& 1 - \mathbb{E}_{L_s}\left[\int_{0}^1vf_I(I^*_{\text{SJD}}(v))\text{d}v\right].\\
\end{align*}
\hrulefill
\end{figure*}
\begin{figure*}
\begin{align*}
&\mathbb{E}_{I}\left[ \max_{0\leq v \leq 1}\min_{0<u\leq 1} \frac{1}{u}\log \left(1+\frac{g(v) - g(v-uv)}{g(1) - g(v) + \frac{I + \sigma^2}{L_s}}\right)\right] \stackrel{(c)}{=}\\
&\mathbb{E}_I\left[\max_{0\leq v \leq 1}\min_{0<u\leq 1}\frac{1}{u}\int_0^{\infty}\frac{e^{-z}}{z}\left(1 - e^{\frac{-z (g(v) - g(v-uv))}{g(1) - g(v) +\frac{I + \sigma^2}{L_s}}}\right)\text{d}z\right]\stackrel{(d)}{=}\\
&\mathbb{E}_I\left[\max_{0\leq v \leq 1}\min_{0<u\leq 1}\frac{1}{u}\int_0^{\infty}\frac{e^{-s((g(1) - g(v))L_s +\sigma^2)}}{s}e^{-sI}\left(1-e^{-s (g(v) - g(v-uv))L_s}\right)\text{d}s\right]\stackrel{(e)}{=}\\
&\max_{0\leq v \leq 1}\min_{0<u\leq 1}\frac{1}{u}\int_0^{\infty}\frac{e^{-s((g(1) - g(v))L_s +\sigma^2)}}{s}\mathbb{E}_I\left[e^{-sI}\right]\left(1-e^{-s (g(v) - g(v-uv))L_s}\right)\text{d}s\stackrel{(f)}{=}\\
&\max_{0\leq v \leq 1}\min_{0<u\leq 1}\frac{1}{u}\int_0^{\infty}\frac{e^{-s((g(1) - g(v))L_s +\sigma^2)}}{s}\psi_I(-s)\left(1-e^{-s (g(v) - g(v-uv))L_s}\right)\text{d}s .
\end{align*}
\hrulefill
\end{figure*}
From (\ref{prop_ordered}) and (\ref{1_e_SD_asym}), the outage probability converges to a deterministic value given below as
\begin{multline*}
\epsilon_{\text{SJD}}(L_s,I,\hat{\textbf{P}}) = 1 - \\ \max_{0\leq v\leq L_s} \left\{v, R \leq  \min_{0 < u \leq 1}\frac{1}{uvL_s}\log \left(1+\frac{g(v)- g(u-uv)}{g(1) - g(v) + \frac{I + \sigma^2}{L_s}}\right)\right\},
\end{multline*}
where $u := \frac{i}{\ell}$ and $v := \frac{\ell}{L_s}$. As we are considering a massive access setting, the average number of singleton layers $\lambda e^{-\frac{\lambda}{L}}$ is taken to be asymptotically large. Moreover, as the ordered received powers converge to deterministic values under massive access, we can see that expression above for the outage probability for a given number of singleton layers depends only on the interference power. Based on this, in step (a), we average over $L_s$ and $I$. Then, in step $(b)$, we decompose the integral such that $I^*_{\text{SJD}}(\frac{\ell}{L_s})$ denotes the maximum interference power for which $\ell$ out of $L_s$ layers can be decoded correctly. It is expressed in (\ref{I_SD}). Finally, we arrive at (\ref{fig_e_SD_asym}) by applying the inversion theorem in (\ref{inv_theorem}).
\section{Proof of Lemma 4}\label{app:4}
The average system throughput can be evaluated by averaging the expression in (\ref{1_zeta_SD_asymp}) over $L_s$ and $I$. Step $(c)$ follows from averaging over the latter and using the following identity [ref]\cite{lin2014downlink}
\begin{equation*}
\log(1 + x) = \int_0^{\infty}\frac{e^{-z}}{z}\left(1-e^{-xz}\right).
\end{equation*}
Step $(d)$ follows from a change of variable: $z = s\left(I + \sigma^2\right)$, and step $(e)$ follows from interchanging the integration and the expectation. Finally, step $(f)$ follows from the definition of CFs.
\section*{Acknowledgement}
The authors would like to thank Dr. Wei Bao and Dr. Matthias Wildemeersch on their valuable feedback on this work.
\bibliographystyle{IEEEtran}
\bibliography{ref}
\end{document}